# New methodology for facilitating food wastage quantification. Identifying gaps and data inconsistencies

Barco Héctor[a, b], Oribe-Garcia Iraia[a,b], Vargas-Viedma, Maria Virginia[a,b], Borges Cruz E[a,b] Martín Cristina[a,b], Alonso-Vicario Ainhoa[a,b]

[a]DeustoTech - Fundación Deusto, Avda. Universidades, 24, 48007, Bilbao, Spain
[b]Facultad Ingeniería, Universidad de Deusto, Avda. Universidades, 24, 48007, Bilbao, Spain

Corresponding Author:
Héctor Barco Cobalea
DeustoTech - Fundación Deusto
Avda. Universidades, 24
48007, Bilbao, Spain

E-mail: hector.barco@deusto.es
Phone: (34) 34.94.413.90.03 ext 3032

# New methodology for facilitating food wastage quantification. Identifying gaps and data inconsistencies

Barco Héctor[a, b, c], Oribe-Garcia Iraia[a,b], Vargas-Viedma, Maria Virginia[a,b], Borges Cruz E[a,b], Martín Cristina[a,b], Alonso-Vicario Ainhoa[a,b]

[a]DeustoTech - Fundación Deusto, Avda. Universidades, 24, 48007, Bilbao, Spain
[b]Facultad Ingeniería, Universidad de Deusto, Avda. Universidades, 24, 48007, Bilbao, Spain
[c]Corresponding autor (hector.barco@deusto.es)

**Highlights**
- Systematic methodologies are key to assess the impacts of food waste generation
- Digital tools for geo-localisation of food waste facilitate the decision making
- Baseline information for food waste reduction is related to economic activities
- Standardisation of methods for quantification helps reduce gaps and inconsistencies

**Abstract**

The Food and Agriculture Organization of the United Nations estimated that about 1.3 billion tons of food produced for human consumption was lost or wasted globally (Gustavsson et al. 2011). Thus, the reduction of the current food loss and waste along the agrifood chain is becoming a priority, both for optimization of resources and reduction waste generation costs. For this purpose, the first step is to quantify the food wastage generation to be able to identify corrective measures. However, in spite of the considerable efforts already undertaken to establish common methodologies to measure the food wastage at different geographical scales, there are still some gaps and inconsistencies.

In this regards, the information gathering is labour-intensive because of the different actors involved. The creation of new methodologies and tools capable of automatically identifying these agents would be of great value so as to subsequently apply the more appropriates quantification methodologies.

This work aims at providing a new methodology to facilitate this process thanks to the previous identification and classification of the potential food wastage generators. As a result, it provides baseline information for one of the earliest steps of the food wastage quantification process, which is the establishment of the scope of the food wastage inventory.

The baseline data needed is taken from the Statistical classification of economic activities in the European Community (NACE), particularly from the most disaggregated level called "classes". This information has been combined with data from the trading income tax at municipal scale thanks to the use of Geographic Information Systems (GIS) and the common codes for NACE classes, generating a visual tool for the localization of points with potential of food-wastage generation and their weight of each economic activity over the agrifood chain.

The proposed methodology has been implemented for the real case of the municipality of Zamudio (Spain) and it has allowed the identification of the different entities linked with economic activities that are potential generators of food wastage, the weight of each activity over the entire agrifood chain, and the geographical location of these entities in the territory. Furthermore, this methodology was used to compare the nature and number of these activities in another municipality (Karrantza, Spain) and it has also been applied to the entire region of the Basque Country (Spain).

## 1. Introduction

According to a report of the Food and Agriculture Organization of the United Nations (FAO), roughly one-third of the food produced for human consumption is lost or wasted globally, which amounts to about 1.3 billion tons per year (Gustavsson et al. 2011). Food wastage occurs at different stages of the food value chain: production, post-production procedures, processing, distribution, and consumption (Lundqvist et al. 2008). According to current estimations (Searchinger et al. 2013; Lipinski et al. 2013), an adequate distribution of calories around the world would require approximately a 63% increase in the demand of crop calories moving from 9,500 trillion kcal per year in 2006 to 15,500 trillion kcal in 2050. Thus, the optimization and a better distribution of the current food resources would be a more suitable solution than the increase of the food production at global level.

However, the negative impacts of the global food waste problem are not only related to the reduced availability and consumption of food but are also directly linked to environmental impacts such as greenhouse gas emissions, consumption of surface and groundwater resources and land occupation (Scialabba et al. 2013). Therefore, thanks to the reduction of the current food loss and waste, it is possible to improve food availability without increasing the agricultural land and environmental impacts. However, a first step to approach corrective and prevention measures for the food waste issue has to be related to the quantification in both industrialized and particularly developing countries because of the lack of food wastage management systems and legislative measures (Thi et al. 2015).

This social and environmental problem has been further emphasized by the European Parliament (2012) and both challenge of food waste measurement and quantification was also addressed within the framework of the European Union (Monier et al. 2011) where the food waste figures at national level were published. These data have served as the starting point for those Member States which have no studies regarding food waste. However, the European Court of Auditors (2016) puts into question the effectiveness provided for by European rules. It includes the target to halve the food waste per capita by 2030 throughout the agrifood chain (United Nations, 2015; European Commission, 2015) because there is no a base year defined in order to set the reduction target for 2030. This lack of information at national level was also expressed in the European project called FUSIONS (Stenmarck et al. 2016) where it is possible to identify significant differences between European Union Member States in terms of the availability of information about food waste at national level.

Despite various studies related to the food wastage at national scale (Reutter et al. 2017; Stensgård et al. 2016; Katajajuuri et al. 2014; Oelofse and Nahman, 2013), there are some gaps and deficiencies at present, in both data and the methodologies used to provide reliable and comparable information about this issue. These gaps and deficiencies cover several aspects such as the lack of a standardized method for these quantification studies, consistent databases or food wastage analyses along the entire agrifood chain (Xue et al. 2017; Chaboud and Daviron, 2017; Bräutigam et al. 2014; Beretta et al. 2013; Partfitt et al. 2010).

Among the methods used to measure the food wastage is possible to define two main approaches (Delgado et al. 2017): the *macro-approach*, using mass and energy balances and *micro-approach*, where there is not any standard method to measure the food waste issue along the entire food chain, but there is a wide variety of methodologies which could be used such as questionnaires, diaries, direct measurements or observations (Møller et al. 2014). Both approaches could use sample data regarding the entire agrifood chain or specific steps (Betz et al. 2015; Quested et al. 2013; Buzby and Hyman, 2012).

A selection of methods, from the macro and micro-approach, was incorporated into the recommendations by FUSIONS on quantitative techniques to estimate the level of wasted food across EU-28 (Møller et al. 2014). Thus, one of the main conclusions was the importance of the harmonization of results between countries, sectors and steps in the food supply chain. For this purpose, FUSIONS recommended the use of the codes from the Statistical classification of economic activities in the European Community (NACE).

The NACE classification features different categories and levels established according to the greater or lesser degree of specificity of the economic activities. FUSIONS suggested use the most specific as possible digit code (Tostivint et al. 2016). These recommendations may represent a step forward in dealing with the current difficulty to compare food waste studies because they are normally adapted to specific objectives, focused only on certain steps along the agrifood chain, and generally using different data and quantification methodologies.

In terms of food wastage, at international level, the Food Loss and Waste Protocol (FLW Protocol) published in 2016 (Hanson et al. 2016) provides a global standard that can be used at national and lower territorial scales (e.g., a region, a province, or a city) and public or private entities. This protocol is ample and flexible to allow the adaptation of all kind of case of studies to measure the food wastage, but it may turn into a certain lack of definition in order to establish strategies for the measurement of food loss and waste in certain territorial scales, such as the local level. This issue can be particularly relevant in one of the first stages for the quantification of the food wastage by the FLW Protocol. This stage is titled *Establishing the scope of an FLW inventory* where it is left to the entities or municipalities the responsibility of defining the points or generating sources of food waste. These points are the places where the measurement of the food wastage subsequently is carried out.

The definition of which entities should become part of the identification of the generating sources of food wastage is one of the earliest steps of the entire protocol. This step is a key aspect to avoid certain gaps and lacks in the definition of the subsequent step of quantifying the food wastage, the measurement itself. The lack of definition at this first stage of the measurement process can eventually result in inactions to tackle the food waste problem.

For this reason, it is necessary to develop methodologies that can support decision-making with regards to the identification of the potentially generating sources of food wastage at local level for their subsequent measurement. Thus, the proposed methodology aims at facilitating decision making processes by defining the location of potential food wastage generators along the entire agrifood chain. Moreover, this methodology uses the NACE codes, seeking the harmonization and the comparability of results which is advocated by FUSIONS, as well as the adaptability of this methodology to different municipalities within the territory of the European Community. In addition, it can be useful to identify possible gaps and inconsistencies at local and supra-local level.

This research work does not differentiate the concept of food loss and food waste according to the FAO definition (FAO, 2014). Thus, we use the concept Food Losses and Waste (FLW) or food wastage to refer to the combined amount of food loss and waste in line with other authors (Xue et al. 2017; Chaboud et al. 2017; Parfit et al. 2011).

## 2. Methodology

The proposed methodology aims at the identification of all the points or generating sources of food waste along the entire agrifood chain at local level so as to facilitate the decision making to the responsible entities on where the measurements regarding the food wastage should be made. It is important to highlight that the methodology proposes to consider the entire agrifood chain, highlighting the following main steps: Production, Manufacturing, Distribution and Retail and Consumption.

Accordingly, the general framework of this methodology is presented in Figure 1:

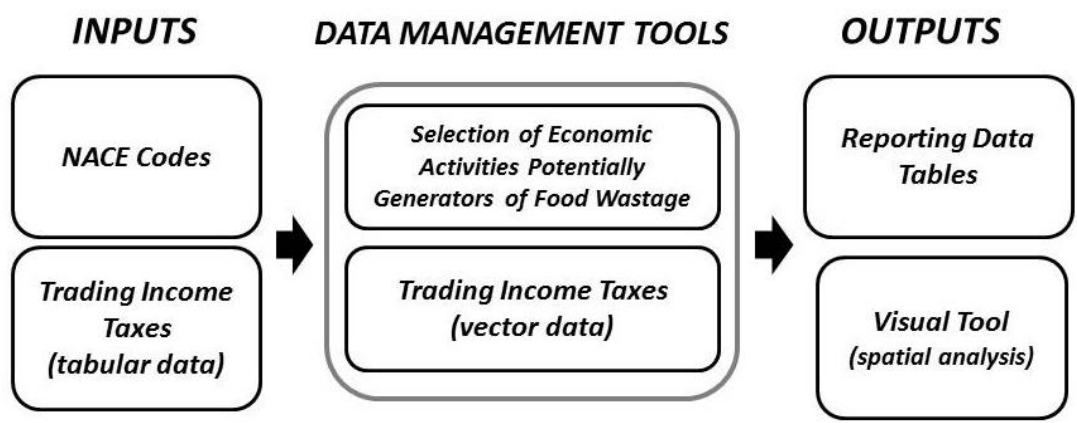

*Figure 1. Theoretical approach on the methodology proposed.*

## 2.1 Inputs

As seen, this methodology has two fundamental inputs:

- Statistical Classification of Economic Activities in the European Community (NACE) codes.
- Trading income taxes.

In the first place, one of these major inputs is the **NACE categories**. NACE is the European standard classification of productive economic activities divided in such a way that a specific code can be associated with an economic activity characterized by an input of resources, a production process and an output of products (goods or services) (European Commission, 2008). This information is freely available and universally accessible. NACE organizes this information in the form of a hierarchical structure where different categories are divided into Sections, Divisions, Groups and Classes (see Table 1):

*Table 1. Example of part of the detailed structure of NACE.*
*(The entire analysis is shown in the Supporting Information).*

| Division Code | Group Code | Class Code | Class Name |
|---|---|---|---|
| colspan=4 | SECTION A — AGRICULTURE, FORESTRY AND FISHING |||
| 01 | | | Crop and animal production, hunting and related service activities |
| | 01,1 | | Growing of non-perennial crops |
| | | 01,11 | Growing of cereals (except rice), leguminous crops and oil seeds |
| | | 01,12 | Growing of rice |
| | | 01,13 | Growing of vegetables and melons, roots and tubers |
| | | 01,14 | Growing of sugar cane |
| | | 01,15 | Growing of tobacco |
| | | 01,16 | Growing of fibre crops |
| | | 01,19 | Growing of other non-perennial crops |
| | 01,2 | | Growing of perennial crops |
| | | 01,21 | Growing of grapes |
| | | 01,22 | Growing of tropical and subtropical fruits |
| | | 01,23 | Growing of citrus fruits |
| | | 01,24 | Growing of pome fruits and stone fruits |
| | | 01,25 | Growing of other tree and bush fruits and nuts |
| | | 01,26 | Growing of oleaginous fruits |
| | | 01,27 | Growing of beverage crops |
| | | 01,28 | Growing of spices, aromatic, drug and pharmaceutical crops |
| | | 01,29 | Growing of other perennial crops |
| | 01,3 | | Plant propagation |
| | | 01,30 | Plant propagation |
| | 01,4 | | Animal production |
| | | 01,41 | Raising of dairy cattle |
| | | 01,42 | Raising of other cattle and buffaloes |
| | | 01,43 | Raising of horses and other equines |
| | | 01,44 | Raising of camels and camelids |
| | | 01,45 | Raising of sheep and goats |

The second main input of the proposed methodology is the **trading income tax** at local level. This is an economic tax which applies to companies carrying out any entrepreneurial or professional activity in a certain territory. The trading income tax contains a specific list of the different entities with economic activity within in a municipality, classified by the NACE categories.

This information can be provided by municipal governments. Thanks to the analysis of the trading income tax, it is possible to generate a list of all entities, public and private, which have an economic activity within a given municipality. In relation to all the information of each entity provided by the trading income tax, it should be highlighted that both the NACE codes and the NACE categories are known as *classes* associated to each entity (see Table 2):

Table 2. Example of part of the trading income taxes.

| Company Name | NACE Class Code | NACE Class Name |
|---|---|---|
| Company 1 | 01.11 | Growing of cereals (except rice), leguminous crops and oil sedes. |
| Company 2 | 01.12 | Growing of rice. |
| Company 3 | 03.12 | Freshwater fishing. |
| Company 4 | 01.14 | Growing of sugar cane. |
| Company 5 | 01.23 | Growing of citrus fruits. |
| Company 6 | 10.84 | Manufacture of condiments and seasonings. |
| Company 7 | 01.50 | Mixed farming. |

2.2 Data Management Tools

Thanks to both inputs, it is possible to create the Data Management Tools that includes two main components:

- Selection of economic activities potentially generators of food wastage.
- Trading income taxes (vector data).

The **selection of economic activities that are potential generators of food wastage** has been created starting from the "NACE Codes" input. The most disaggregated NACE category (class) is considered. Thus, the selection of these categories keeps in line with the main conclusions drawn by FUSIONS with regard to using the more specific as possible digit code (Tostivint et al. 2016).

In order to determine if the economic activity included in each class may be susceptible to generate food wastage, a comprehensive analysis of each and every single-class category has been carried out thanks to the use of expert criteria and by analyzing the definition of each of the classes which compose the complete structure of NACE.

Considering the classes proposed by NACE, three types of typologies have been distinguished:

- **Potential Food Wastage Generation.** Commercial activities defined by the official EUROSTAT document (European Commission, 2008) that might be susceptible to generate food wastage.
- **Non-Potential Food Wastage Generation**. Commercial activities that cannot be susceptible to generate food wastage.
- **In-situ Verification**. The case of a group of commercial activities where some of these may occasionally produce food wastage and others that are catalogued as non-potential food wastage generators. For this reason, it is necessary to verify in-situ the specific economic activity linked with the specific case and its relationship with the agrifood value chain.

Table 3 shows an example of this analysis which comprises the entire NACE code, where an identification of each of the NACE classes in these three different typologies has been carried out:

*Table 3. Example of part of the detailed structure of NACE, identifying the different types of classes in accordance with their potentiality as a food wastage generator. (The entire analysis is shown in the Supporting Information).*

| Division Code | Group Code | Class Code | Class Name | Food Waste Code |
|---|---|---|---|---|
| colspan=5 | SECTION A — AGRICULTURE, FORESTRY AND FISHING | | | |
| 01 | | | Crop and animal production, hunting and related service activities | |
| | 01,1 | | Growing of non-perennial crops | |
| | | 01,11 | Growing of cereals (except rice), leguminous crops and oil seeds | PFW |
| | | 01,12 | Growing of rice | PFW |
| | | 01,13 | Growing of vegetables and melons, roots and tubers | PFW |
| | | 01,14 | Growing of sugar cane | PFW |
| | | 01,15 | Growing of tobacco | NPFW |
| | | 01,16 | Growing of fibre crops | NPFW |
| | | 01,19 | Growing of other non-perennial crops | NPFW |
| | 01,2 | | Growing of perennial crops | |
| | | 01,21 | Growing of grapes | PFW |
| | | 01,22 | Growing of tropical and subtropical fruits | PFW |
| | | 01,23 | Growing of citrus fruits | PFW |
| | | 01,24 | Growing of pome fruits and stone fruits | PFW |
| | | 01,25 | Growing of other tree and bush fruits and nuts | PFW |
| | | 01,26 | Growing of oleaginous fruits | PFW |
| | | 01,27 | Growing of beverage crops | PFW |
| | | 01,28 | Growing of spices, aromatic, drug and pharmaceutical crops | PFW |
| | | 01,29 | Growing of other perennial crops | NPFW |
| | 01,3 | | Plant propagation | |
| | | 01,30 | Plant propagation | NPFW |
| | 01,4 | | Animal production | |
| | | 01,41 | Raising of dairy cattle | PFW |
| | | 01,42 | Raising of other cattle and buffaloes | PFW |
| | | 01,43 | Raising of horses and other equines | PFW |
| | | 01,44 | Raising of camels and camelids | PFW |
| | | 01,45 | Raising of sheep and goats | PFW |
| | | 01,46 | Raising of swine/pigs | PFW |
| | | 01,47 | Raising of poultry | PFW |
| | | 01,49 | Raising of other animals | IV |
| | 01,5 | | Mixed farming | |
| | | 01,50 | Mixed farming | PFW |

**Food Waste Codes:**
Potential Food Wastage Generation = PFW
No Potential Food Wastage Generation = NPFW
In-Situ Verification = IV

In this regard, the collection of data from the upper level categories such as the group *01,1 - Growing of non-perennials crops* has some inconsistencies because despite the fact that it includes economic activities related to potential food wastage generators, as is the case with the class *01,12 - Growing of rice*, it also includes other economic activities which are not linked with the potential generation of food wastage, as is the case with the class *01,15 - Growing of tobacco*. Thus, it could be possible to highlight the importance of establishing food waste analysis by NACE classes instead of only using upper level categories.

After defining all classes from NACE, under one of the three types of typologies stated previously, those categories which are referred to as Potential Food Wastage Generation and In-situ Verification have been selected and classified according to the stage of the agrifood chain to which they belong to, as can be seen in Table 4:

*Table 4. Example of part of the detailed structure of economic activities in accordance with their potentiality as a food wastage generator categorized by steps of the agrifood chain and NACE codes. (The entire analysis is shown in the Supporting Information).*

| Step Agrifood Chain | Division Code | Group Code | Class Code | Class Name | Food Waste Code |
|---|---|---|---|---|---|
| | | | | **SECTION A — AGRICULTURE, FORESTRY AND FISHING** | |
| | 01 | | | **Crop and animal production, hunting and related service activities** | |
| | | 01,1 | | Growing of non-perennial crops | |
| | | | 01,11 | Growing of cereals (except rice), leguminous crops and oil seeds | PFW |
| | | | 01,12 | Growing of rice | PFW |
| | | | 01,13 | Growing of vegetables and melons, roots and tubers | PFW |
| | | | 01,14 | Growing of sugar cane | PFW |
| | | 01,2 | | Growing of perennial crops | |
| | | | 01,21 | Growing of grapes | PFW |
| | | | 01,22 | Growing of tropical and subtropical fruits | PFW |
| | | | 01,23 | Growing of citrus fruits | PFW |
| | | | 01,24 | Growing of pome fruits and stone fruits | PFW |
| **PRODUCTION** | | | 01,25 | Growing of other tree and bush fruits and nuts | PFW |
| | | | 01,26 | Growing of oleaginous fruits | PFW |
| | | | 01,27 | Growing of beverage crops | PFW |
| | | | 01,28 | Growing of spices, aromatic, drug and pharmaceutical crops | PFW |
| | | 01,4 | | Animal production | |
| | | | 01,41 | Raising of dairy cattle | PFW |
| | | | 01,42 | Raising of other cattle and buffaloes | PFW |
| | | | 01,43 | Raising of horses and other equines | PFW |
| | | | 01,44 | Raising of camels and camelids | PFW |
| | | | 01,45 | Raising of sheep and goats | PFW |
| | | | 01,46 | Raising of swine/pigs | PFW |
| | | | 01,47 | Raising of poultry | PFW |
| | | | 01,49 | Raising of other animals | IV |
| | | 01,5 | | Mixed farming | |
| | | | 01,50 | Mixed farming | PFW |

**Food Waste Codes:**
Potential Food Wastage Generation = PFW
In-Situ Verification = IV

The second element of the Data Management Tools is the ***Trading income taxes (vector data)***. The main advantage of the collection of trading income tax for this methodology is the possibility of obtaining a list of all the entities in the municipal area with an economic activity and their linkage with a particular class from NACE, identified by a specific code. The proposed methodology takes as starting point the use of the trading income taxes from the municipality of Zamudio, as it is the case of study of the research work.

Thanks to the use of the Geographic Information Systems (GIS), and particularly of the use of the software ArcGIS 10.6, the entire group of the business registers in the city Zamudio has been geo-positioned through the assignment of geographic coordinates (Latitude and Longitude) included in the trading income tax for later conversion back to vector data, using the point typology and referred to the WGS-84 datum (World Geodetic System 1984). In cases where these geocodes are not available in the trading income tax, it is possible to obtain those thanks to the postal address information of each of the entities because these data are included in the trading income tax. As shown in Tables 2 and 5, it is possible to observe how the available information from the trading income tax is transformed from tabular data into vector data. Thanks to the geographic coordinates, it allows the geolocation of the economic activities with potential generation of food waste at municipal level.

*Table 5. Example of the trading income tax transformed from tabular data to vector data.*

| FID | Shape | Company Name | Class Code | NACE Class Name | Latitude | Longitude |
|---|---|---|---|---|---|---|
| 1 | Point | Company 1 | 01.11 | Growing of cereals (except rice), leguminous crops and | 43,277946 | -2,871942 |
| 2 | Point | Company 2 | 01.12 | Growing of rice. | 42,234911 | -2,862493 |
| 3 | Point | Company 3 | 03.12 | Freshwater fishing. | 42,214503 | -2,852589 |
| 4 | Point | Company 4 | 01.14 | Growing of sugar cane. | 42,293813 | -2,862562 |
| 5 | Point | Company 5 | 01.23 | Growing of citrus fruits. | 42,281283 | -2,864891 |
| 6 | Point | Company 6 | 10.84 | Manufacture of condiments and seasonings. | 42,249912 | -2,859322 |
| 7 | Point | Company 7 | 01.50 | Mixed farming. | 42,289103 | -2,875502 |

In this way, by transforming the information obtained from the trading income tax and the inclusion of the geographical coordinates, if such data were not available, it is possible to merge those with the selection of economic activities that are potential generators of food wastage, mentioned above. To do this, thanks to the use of the above-mentioned GIS software, the merging of data is normally used to incorporate fields of a data table with the other through the use of the fields with common attributes to both data tables. In this case, the common field to merge the trading income taxes and the economic activities potentially generators of food wastage has been the NACE code. On the basis of this merging, a new vector layer composed of points (shapefile) has been obtained with the group of the selection of the entities linked to economic activities potentially generators of food wastage (Potential Food Wastage Generation and In-Situ Verification) in the municipality of Zamudio.

Thanks to the use of the GIS, it is possible to link the information on typologies classified as Potential Food Wastage Generation, Non-Potential Food Wastage Generation and In-situ Verification, including each of the entities belonging to the municipality studied. Thus, the element related to selection of economic activities is used as a filter to obtain two main outputs from this methodology which are explained in Section 3.

### 2.3 Outputs

The most relevant results thanks to the use of this methodology are the following outputs:

- Reporting Data Tables.
- Visual Tool (spatial analysis)

Both outputs provide the baseline information for the first step of the food wastage quantification process. The reporting data tables gather the information related to the nature of the entities linked with economic activities with potential generation of food wastage and their weight of each activity over the entire agrifood chain at local scale. The visual tool is focused on the geographical location of these entities susceptible to produce food wastage on the territory. The main outputs from this methodology are explained in the following section.

### 3. Results

The proposed methodology has been implemented for the real case of the municipality of Zamudio (Spain) and it has enabled the identification of the different entities linked with economic activities that are potential generators of food wastage, the weight of each activity over the entire agrifood chain, and the geographical location of these entities in the territory.

Furthermore, in order to demonstrate its replication capabilities throughout the European Union at the municipal scale, this methodology was used to compare the nature and number of these activities in another municipality (Karrantza, Spain) with similar population size analyzing their similarities and differences regarding the food wastage issue. In addition, this method has also been applied to the region of the Basque Country in order to analyze its possibilities of adapting to larger territorial contexts such as the regional and national scales, detecting possible hotspots and points of improvement of the current official figures at this respect.

### 3.1 The municipality of Zamudio

Zamudio is an industrial city located in Spain, within the region of the Basque Country and it has a population of 2703 inhabitants.

The information of the different economic activities in Zamudio that could potentially generate food wastage are displayed through the **data tables**. These data tables reflect the entities defined as Potential Food Wastage Generation and In-situ Verification as well as the number of these entities with respect to their economic activities linked to their potential for the generation of food wastage.

*Table 6. Case of study of the municipality of Zamudio. Number of entities with potential for food wastage generation categorized by steps of the agrifood chain and sections, divisions, groups and classes categories from NACE.*

| Step Agrifood Chain | Division Code | Group Code | Class Code | Class Name | Number of entities |
|---|---|---|---|---|---|
| | | | | **SECTION A — AGRICULTURE, FORESTRY AND FISHING** | |
| **PRODUCTION** | **01** | | | **Crop and animal production, hunting and related service activities** | |
| | | 01,4 | | Animal production | |
| | | | 01,49 | Raising of other animals | 1 |
| | | | | **SECTION C — MANUFACTURING** | |
| | **10** | | | **Manufacture of food products** | |
| | | 10,1 | | Processing and preserving of meat and production of meat products | |
| | | | 10,13 | Production of meat and poultry meat products | 1 |
| | | 10,2 | | Processing and preserving of fish, crustaceans and molluscs | |
| | | | 10,20 | Processing and preserving of fish, crustaceans and molluscs | 1 |
| | | 10,5 | | Manufacture of dairy products | |
| **MANUFACTURING** | | | 10,51 | Operation of dairies and cheese making | 2 |
| | | 10,8 | | Manufacture of other food products | |
| | | | 10,83 | Processing of tea and coffee | 1 |
| | **11** | | | **Manufacture of beverages** | |
| | | 11,0 | | Manufacture of beverages | |
| | | | 11,02 | Manufacture of wine from grape | 1 |
| | | | 11,03 | Manufacture of cider and other fruit wines | 1 |
| | | | 11,05 | Manufacture of beer | |
| | | | | **SECTION G — WHOLESALE AND RETAIL TRADE; REPAIR OF MOTOR VEHICLES AND MOTORCYCLES** | |
| | **46** | | | **Wholesale trade, except of motor vehicles and motorcycles** | |
| | | 46,3 | | Wholesale of food, beverages and tobacco | |
| | | | 46,31 | Wholesale of fruit and vegetables | 1 |
| | | | 46,32 | Wholesale of meat and meat products | 4 |
| | | | 46,33 | Wholesale of dairy products, eggs and edible oils and fats | 2 |
| | | | 46,34 | Wholesale of beverages | 7 |
| | | | 46,36 | Wholesale of sugar and chocolate and sugar confectionery | 2 |
| | | | 46,37 | Wholesale of coffee, tea, cocoa and spices | 1 |
| | | | 46,38 | Wholesale of other food, including fish, crustaceans and molluscs | 1 |
| | | | 46,39 | Non-specialised wholesale of food, beverages and tobacco | 6 |
| **DISTRIBUTION AND RETAIL** | **47** | | | **Retail trade, except of motor vehicles and motorcycles** | |
| | | 47,1 | | Retail sale in non-specialised stores | |
| | | | 47,11 | Retail sale in non-specialised stores with food, beverages or tobacco | 2 |
| | | 47,2 | | Retail sale of food, beverages and tobacco in specialised stores | |
| | | | 47,22 | Retail sale of meat and meat products in specialised stores | 2 |
| | | | 47,23 | Retail sale of fish, crustaceans and molluscs in specialised stores | 1 |
| | | | 47,24 | Retail sale of bread, cakes, flour confectionery and sugar confectionery in | 3 |
| | | | 47,29 | Other retail sale of food in specialised stores | 1 |
| | | | | **SECTION H — TRANSPORTATION AND STORAGE** | |
| | **49** | | | **Land transport and transport via pipelines** | |
| | | 49,4 | | Freight transport by road and removal services | |
| | | | 49,41 | Freight transport by road | 2 |
| | | | | **SECTION I — ACCOMMODATION AND FOOD SERVICE ACTIVITIES** | |
| | **55** | | | **Accommodation** | |
| | | 55,1 | | Hotels and similar accommodation | |
| | | | 55,10 | Hotels and similar accommodation | 2 |
| | | 55,2 | | Holiday and other short-stay accommodation | |
| | | | 55,20 | Holiday and other short-stay accommodation | 1 |
| | | 56,1 | | Restaurants and mobile food service activities | |
| | | | 56,10 | Restaurants and mobile food service activities | 13 |
| | | 56,2 | | Event catering and other food service activities | |
| | | | 56,21 | Event catering activities | 1 |
| | | | 56,29 | Other food service activities | 2 |
| | | 56,3 | | Beverage serving activities | |
| | | | 56,30 | Beverage serving activities | 9 |
| | | | | **SECTION P — EDUCATION** | |
| | **85** | | | **Education** | |
| | | 85,1 | | Pre-primary education | |
| | | | 85,10 | Pre-primary education | 1 |
| | | 85,2 | | Primary education | |
| **CONSUMPTION** | | | 85,20 | Primary education | 1 |
| | | | | **SECTION Q — HUMAN HEALTH AND SOCIAL WORK ACTIVITIES** | |
| | **86** | | | **Human health activities** | |
| | | 86,1 | | Hospital activities | |
| | | | 86,10 | Hospital activities | 1 |
| | | 86,9 | | Other human health activities | |
| | | | 86,90 | Other human health activities | 1 |
| | **87** | | | **Residential care activities** | |
| | | 87,3 | | Residential care activities for the elderly and disabled | |
| | | | 87,30 | Residential care activities for the elderly and disabled | 1 |
| | | 88,9 | | Other social work activities without accommodation | |
| | | | 88,91 | Child day-care activities | 1 |
| | | | | **SECTION R — ARTS, ENTERTAINMENT AND RECREATION** | |
| | **93** | | | **Sports activities and amusement and recreation activities** | |
| | | 93,1 | | Sports activities | |
| | | | 93,12 | Activities of sport clubs | 2 |
| | | 93,2 | | Amusement and recreation activities | |
| | | | 93,29 | Other amusement and recreation activities | 2 |

Table 6 presents the number of entities that are potential food wastage generators along the different stages of the agrifood chain in the municipality of Zamudio. In fact, there is a total of 82 entities with potential food wastage generation with 4 of them belonging In situ Verification category. If the entities are analyzed by the steps of the agrifood chain, it is possible to appreciate the condition of Zamudio as an industrial city because the majority of its economic activities, at least those entities which could generate food wastage, are located in the Distribution and Consumption level and very few in the Production step. (See Figure 2).

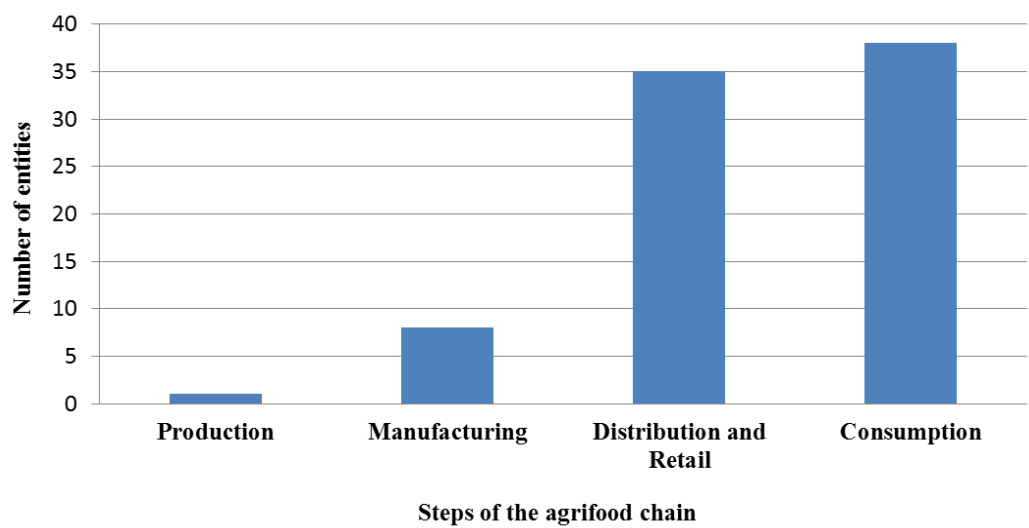

*Figure 2. Case of study of the municipality of Zamudio. Number of entities with potential for food wastage generation categorized by steps of the agrifood chain.*

There is only one economic activity related to the Production step of the entire agrifood chain which is a potential generator of food wastage in Zamudio. This entity is linked with the *Raising of other animals category* and considered as In-situ Verification. The Manufacturing step in the municipality of Zamudio includes 8 entities with potential food waste generation, 5 of them are linked with the *Manufacture of food products* Division and the remaining three are within the *Manufacture of Beverages* Division.

By analyzing the different divisions, it is possible to define which economic activities are associated to entities with potential food wastage generation (See Figure 3).

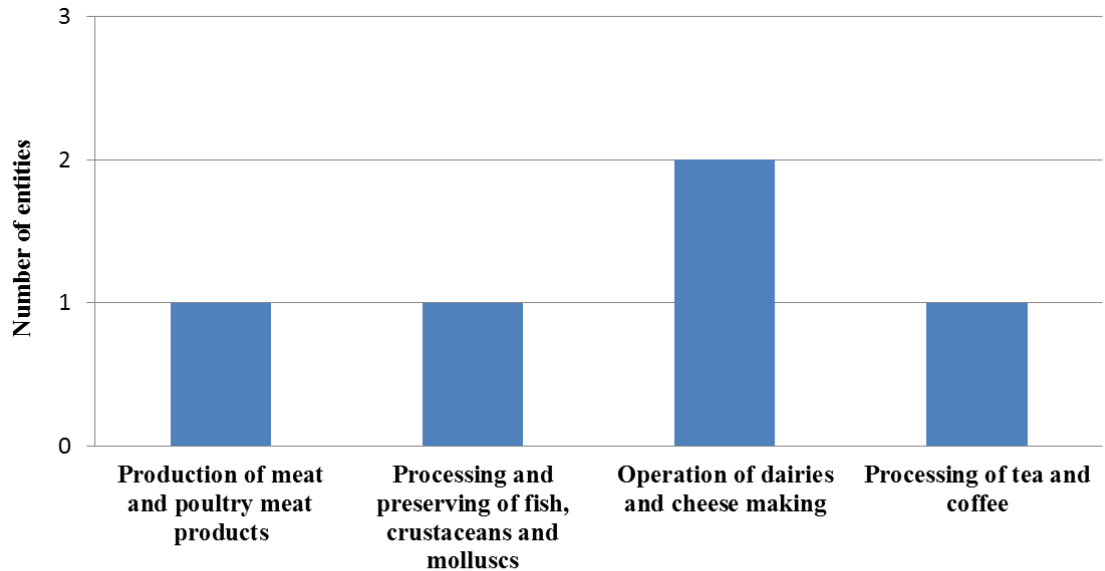

*Figure 3. Case of study of the municipality of Zamudio. Number of entities with potential for food wastage generation classified into different NACE classes within the Division "Manufacture of food products".*

Figure 3 illustrates an example in the municipality of Zamudio. Within the *Manufacture of food products* Division, it is possible to define the different economic activities, using the Class category from NACE, which are linked with the entities that are potential generators of food wastage in Zamudio. These analyses are useful to provide further information about the different entities with potential food wastage generation and the possible type of food loss and waste that they could generate. In the case of Zamudio there is a very similar percentage of manufacturing entities producing meat, fish, dairies and beverages.

Regarding the distribution and retail sector in Zamudio, there are 35 entities with potential food wastage generation, 2 of them categorized as In-Situ Verification. 24 of 35 entities are wholesale trade, 9 retail trade and 2 related to the transport and storage sector.

It is also possible to analyze the different economic activities within the **wholesale trade** with potential food wastage generation in Zamudio for a more detailed account of the nature of the economics activities within this Division (see Figure 4).

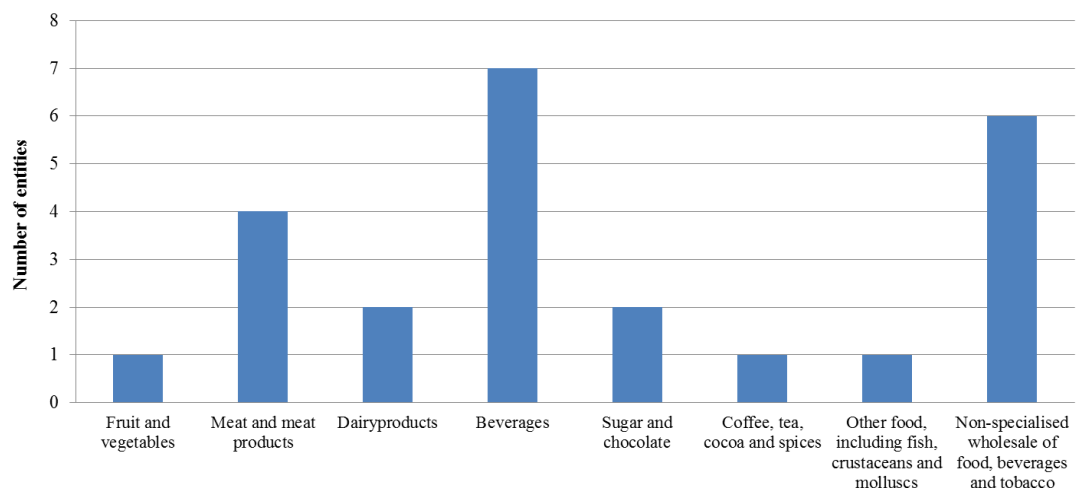

*Figure 4. Case of study of the municipality of Zamudio. Number of entities with potential for food wastage generation classified into different NACE classes within the Division "Wholesale trade".*

Related to the consumption step, it is necessary to highlight that households are excluded from these analysis because they are not considered as economic activities. Despite the fact that the food wastage in households are relevant in the territory (Quested et al. 2011) and that there is little authoritative data on food waste quantities, composition and systematic and comparable information (Blas et al. 2018; Lebersorger and Schneider, 2011), the identification and categorization of work in households at local scale in order to help define which points are more relevant to the subsequent food wastage measurement are out of the scope of this research and will be an interesting issue for future activities.

The case of the municipality of Zamudio includes 38 entities with potential food wastage generation into the Consumption step (and excluding households), 3 of them are considered as In-Situ Verification. 28 of 38 entities are within the *Accommodation and Food Services Activities* Section, 2 in the Education Section, 4 in the Human Health and Social Work Activities Section and 4 in Arts, Entertainment and Recreation.

Thus, the majority of entities with potential food wastage generation, more than 73%, are included in the *Accommodation and Food Services Activities* Section. For this reason, it would be the most important section in which to analyze the different economic activities linked with these 28 entities (see Figure 5).

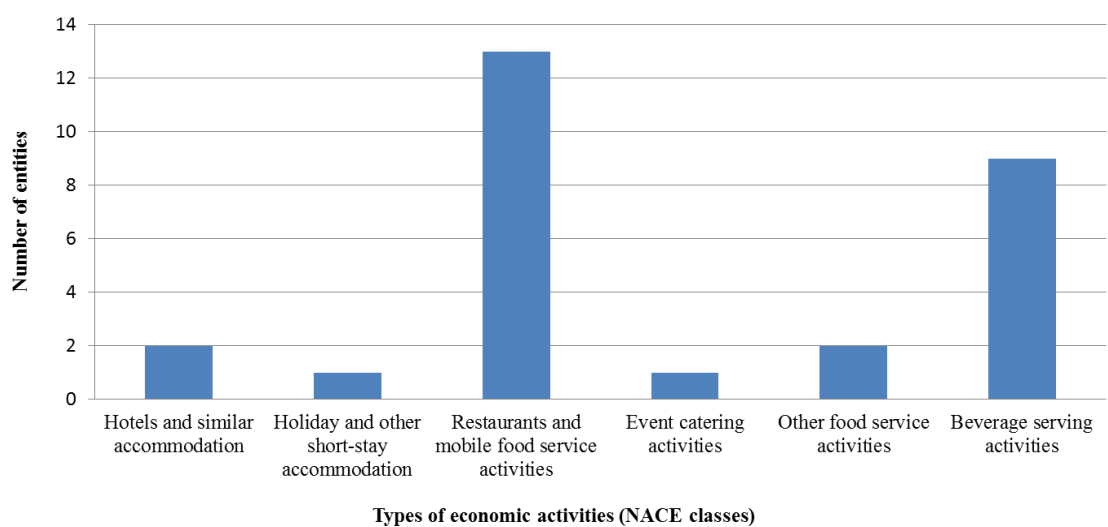

*Figure 5. Case of study of the municipality of Zamudio. Number of entities with potential food wastage generation classified into different NACE classes within the Section "Accommodation and Food Services Activities".*

Despite the fact that *Accommodation and Food Services Activities* Section covers a large amount of economic activities, Figure 5 shows that for the case of Zamudio, two economic activities *Restaurants and mobile food service activities* with 13 entities and *Beverage serving activities* with 9 entities stand among the others significantly. This information and analyses could help to optimize the decision making process to define subsequent food wastage measurements at local level.

3.2 <u>Comparison with the municipality of Karrantza.</u>

Another outcome obtained from the Reporting Data Tables is the possibility of generating a comparative framework between different municipalities around the European Community. Thus, it delimits all the economic activities likely to generate food surpluses for each stage of the agrifood chain in different municipalities.

As shown in the Table 7, the results from the municipality of Zamudio have been compared with those of the municipality of Karrantza. This municipality is also located in Spain and the Basque Country region and has a population of 2.729 inhabitants. Thus, Karrantza has a similar population size but with a clear agricultural character, Table 7 shows the differences between economic activities with potential generation of food wastage in the production level, where the municipality of Zamudio only has an entity and Karrantza boasts a total of 180 entities.

*Table 7. Case of study of the comparison between the municipalities of Zamudio and Karrantza. Part of the number of entities with potential food wastage generation categorized by steps of the agrifood chain and NACE codes. (Further details are shown in the Supporting Information)*

| Step Agrifood Chain | Division Code | Group Code | Class Code | Class Name | Number of entities | |
|---|---|---|---|---|---|---|
| | | | | | Zamudio | Karrantza |
| | | | | **SECTION A — AGRICULTURE, FORESTRY AND FISHING** | | |
| P R O D U C T I O N | **01** | | | **Crop and animal production, hunting and related service activities** | | |
| | | 01,2 | | Growing of perennial crops | | |
| | | | 01,25 | Growing of other tree and bush fruits and nuts | | 1 |
| | | 01,4 | | Animal production | | |
| | | | 01,41 | Raising of dairy cattle | | 87 |
| | | | 01,42 | Raising of other cattle and buffaloes | | 50 |
| | | | 01,43 | Raising of horses and other equines | | 2 |
| | | | 01,45 | Raising of sheep and goats | | 14 |
| | | | 01,47 | Raising of poultry | | 1 |
| | | | 01,49 | Raising of other animals | 1 | 19 |
| | | 01,5 | | Mixed farming | | |
| | | | 01,50 | Mixed farming | | 4 |
| | | 01,6 | | Support activities to agriculture and post-harvest crop activities | | |
| | | | 01,61 | Support activities for crop production | | 1 |
| | | 01,7 | | Hunting, trapping and related service activities | | |
| | | | 01,70 | Hunting, trapping and related service activities | | 1 |
| | | | | **SECTION C — MANUFACTURING** | | |
| M A N U F A C T U R I N G | **10** | | | **Manufacture of food products** | | |
| | | 10,1 | | Processing and preserving of meat and production of meat products | | |
| | | | 10,11 | Processing and preserving of meat | | 1 |
| | | | 10,13 | Production of meat and poultry meat products | 1 | 1 |
| | | 10,2 | | Processing and preserving of fish, crustaceans and molluscs | | |
| | | | 10,20 | Processing and preserving of fish, crustaceans and molluscs | 1 | |
| | | 10,5 | | Manufacture of dairy products | | |
| | | | 10,51 | Operation of dairies and cheese making | 2 | |
| | | | 10,52 | Manufacture of ice cream | | 1 |
| | | 10,6 | | Manufacture of grain mill products, starches and starch products | | |
| | | | 10,61 | Manufacture of grain mill products | | 1 |
| | | 10,8 | | Manufacture of other food products | | |
| | | | 10,83 | Processing of tea and coffee | 1 | |
| | **11** | | | **Manufacture of beverages** | | |
| | | 11,0 | | Manufacture of beverages | | |
| | | | 11,02 | Manufacture of wine from grape | 1 | |
| | | | 11,03 | Manufacture of cider and other fruit wines | 1 | |
| | | | 11,05 | Manufacture of beer | 1 | |

Furthermore, it is possible to compare the total number of entities with potential of food wastage generation along the different steps of the agrifood chain (see Figure 6).

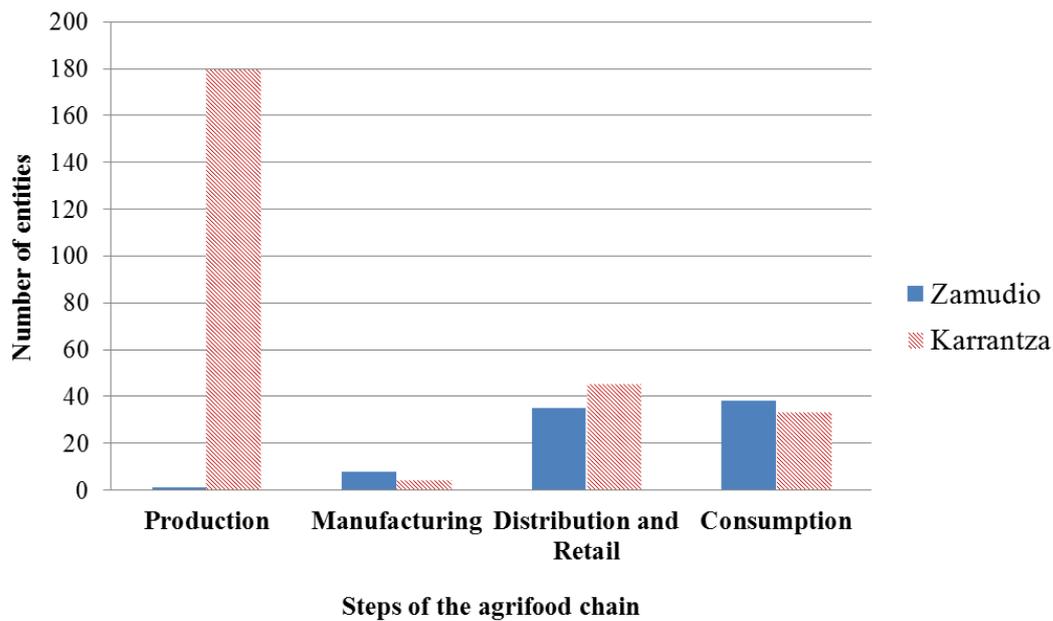

*Figure 6. Case of study of the comparison between the municipalities of Zamudio and Karrantza. Total number of entities with potential food wastage generation categorized by steps of the agrifood chain.*

This type of comparative analyses has been conducted on different levels (Sections, Divisions, Groups or Classes) because sometimes it may happen that both municipalities have similar number of entities with potential generation of food wastage in a specific step of the agrifood chain but the nature of economic activities involved is different (See Figure 7).

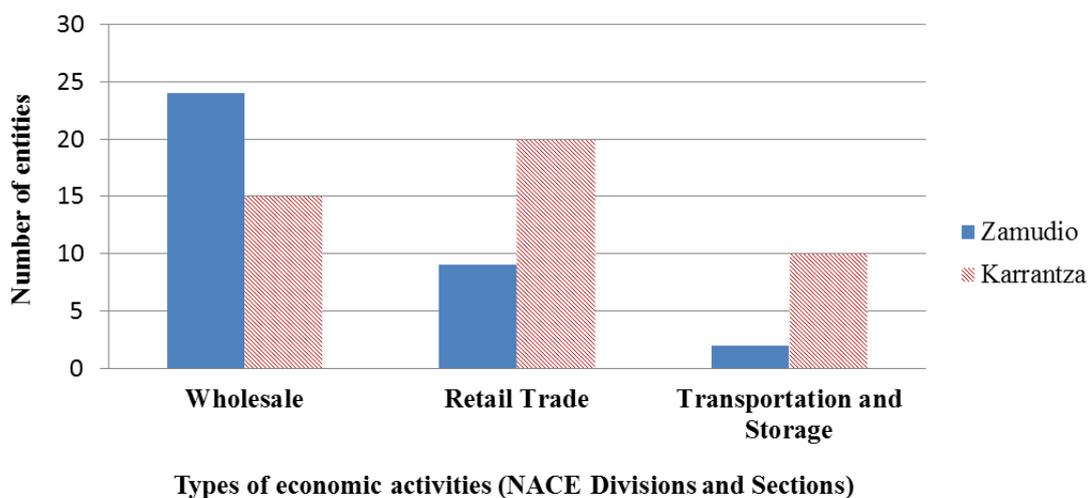

*Figure 7. Case of study of the comparison between the municipalities of Zamudio and Karrantza. Total number of entities with potential food wastage generation in the distribution and retail step.*

Figure 7 shows an example of this issue because despite the total number of entities with potential generation of food wastage within the distribution and retail sector is very similar between the municipalities of Zamudio and Karrantza, 35 and 45 respectively, the differences between wholesale and retail trade entities are relevant. The municipality of Zamudio has a numeric dominance with the wholesales with respect to retail trades and the municipality of Karrantza has fewer differences, in fact there are much more retail trades. These differences could be increased if the analysis is carried out within the different types of wholesale and retail trade activities in Zamudio and Karrantza.

Therefore, it might be concluded that the strategies in Zamudio and Karrantza to measure the food wastage and to find ways of reducing the food wastage issue would be different in certain aspects because of the significant differences in the economic activities which take place in both municipalities.

The second input from this methodology is the **visual map**, where all potential focus of food wastage generation are defined along the entire agrifood chain and classified according to the different stages or NACE categories.

Thus, the visual tool is a point map showing the potentially generating sources of food waste at local level for their subsequent measurements.

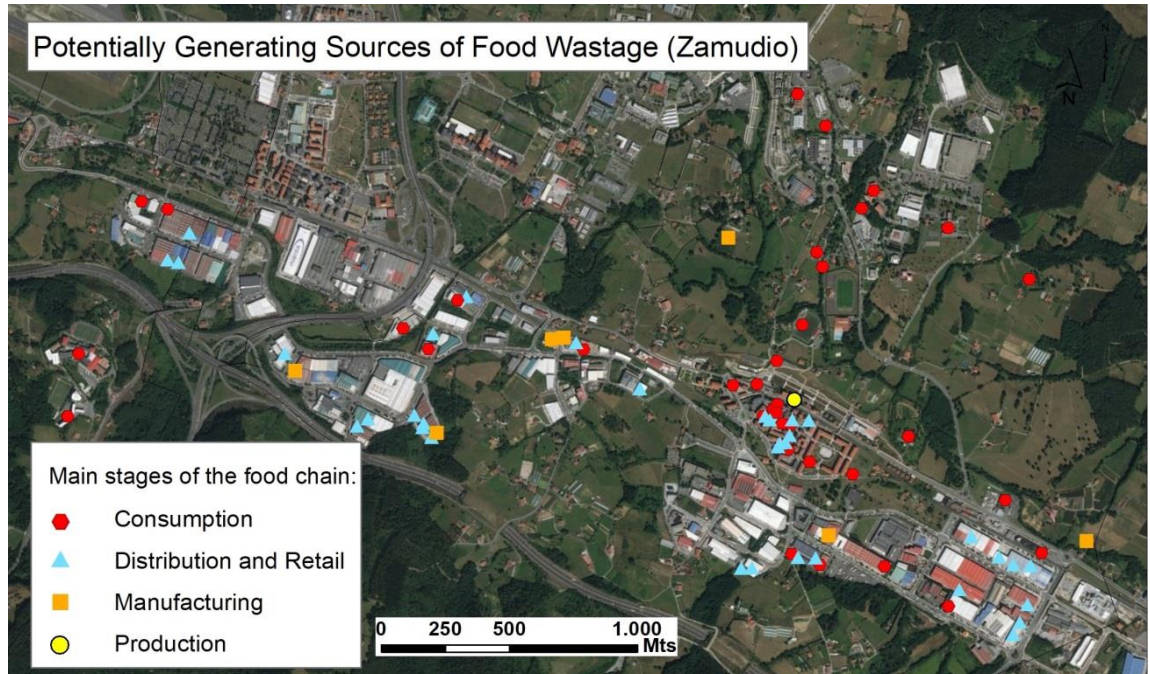

*Figure 8. Case of study of the Localization of Potential Food Wastage Generators.*
*The municipality of Zamudio.*

Figure 8 shows the geographical location of all entities susceptible to quantify the food wastage situation disaggregated by the four main steps of the agrifood chain in Zamudio. Similarly, this break-down into the steps of the agrifood chain where the points are categorized is also possible to disaggregate into the different categories used in the NACE codes, related to their economic activity: Section, Division, Group and Class. Furthermore, obtaining additional comparative information between different municipalities is possible thanks to the use of the visual tool.

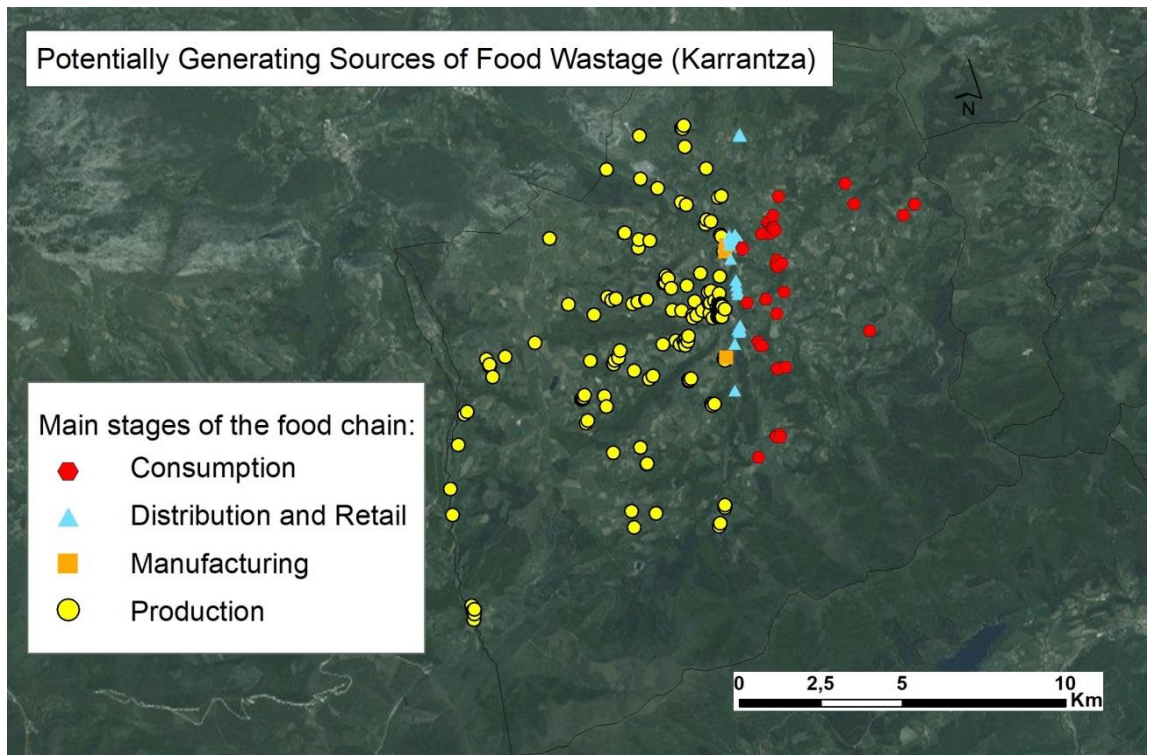

*Figure 9. Case of study of the Localization of Potential Food Wastage Generators. The municipality of Karrantza.*

When comparing Figure 8 and 9, it is possible to observe the disparities of geographical scales between both maps. Thus, Figure 8 shows an image of the urban area and all the entities susceptible to generate food waste within these limits. On the contrary, Figure 9 plots the entire municipal territory. The main explanation is related to the number of entities potentially generators of food wastage linked with the production step. The municipality of Karrantza boasts a total of 180 entities and Zamudio only 1 and the nature of these economic activities does not allow being exclusively included in the urban area. This information should be also considered in order to create strategies to the measurement process and the subsequent corrective measures in both municipalities.

Thanks to the information provided by the Reporting Data Tables, related to the number of entities linked with economic activities considered as potential for food wastage generation and the nature of these economical activities classified according to the different NACE categories, and also thanks to the spatial analysis of these entities facilitated by the use of the visual tool, any agent could create a baseline information in order to facilitate the decision making, policy definition, establishment of strategies, etc. for the quantification of the food loss and waste, and it could contribute to fill the gap detected in the first stages for the quantification of the food loss and waste by the FLW Protocol to be applied at local level.

It is considered a key aspect to create baseline information so as to define the scope of an FLW inventory, particularly in some territorial scales such as the local level. In other words, the baseline information provided by the current methodology aims to help answer where food wastage should be measured. The definition of the scope of an FLW inventory is the earlier stage of the measurement methods to quantify properly the food wastage in the territory.

Additionally, it is important to underline that the food waste quantification is the first step to achieve the main objective related to the reduction of the current food waste. In fact, despite the fact that there are modern alternatives for the food wastage treatment, the food wastage prevention represents greater environmental benefits (Bernstad Saraiva Schott and Andersson, 2015).

   3.3 <u>The region of the Basque country.</u>

Finally, with regard to the applicability of the proposal methodology, it is worth noting that the flexibility of the methodology is not only related to different characteristics of municipalities across the European

Community but it is also possible to adapt to other territorial scales. Table 8 identifies the economic activities that potentially generate food wastage at regional level.

*Table 8. Case of study of the Basque Country (Spain). Part of the number of entities with potential for food wastage generation categorized by steps of the agrifood chain and sections, divisions, groups and classes categories from NACE. (further details are shown in the Supporting Information).*

| Step Agrifood Chain | Division Code | Group Code | Class Code | Class Name | Number of entities |
|---|---|---|---|---|---|
| PRODUCTION | | | | **SECTION A — AGRICULTURE, FORESTRY AND FISHING** | |
| | 01 | | | **Crop and animal production, hunting and related service activities** | |
| | | 01,1 | | Growing of non-perennial crops | |
| | | | 01,11 | Growing of cereals (except rice), leguminous crops and oil seeds | 521 |
| | | | 01,12 | Growing of rice | 0 |
| | | | 01,13 | Growing of vegetables and melons, roots and tubers | 452 |
| | | | 01,14 | Growing of sugar cane | 0 |
| | | 01,2 | | Growing of perennial crops | |
| | | | 01,21 | Growing of grapes | 694 |
| | | | 01,22 | Growing of tropical and subtropical fruits | 11 |
| | | | 01,23 | Growing of citrus fruits | 0 |
| | | | 01,24 | Growing of pome fruits and stone fruits | 53 |
| | | | 01,25 | Growing of other tree and bush fruits and nuts | 10 |
| | | | 01,26 | Growing of oleaginous fruits | 10 |
| | | | 01,27 | Growing of beverage crops | 0 |
| | | | 01,28 | Growing of spices, aromatic, drug and pharmaceutical crops | 0 |
| | | 01,4 | | Animal production | |
| | | | 01,41 | Raising of dairy cattle | 476 |
| | | | 01,42 | Raising of other cattle and buffaloes | 865 |
| | | | 01,43 | Raising of horses and other equines | 12 |
| | | | 01,44 | Raising of camels and camelids | 0 |
| | | | 01,45 | Raising of sheep and goats | 440 |
| | | | 01,46 | Raising of swine/pigs | 18 |
| | | | 01,47 | Raising of poultry | 93 |
| | | | 01,49 | Raising of other animals | 386 |
| | | 01,5 | | Mixed farming | |
| | | | 01,50 | Mixed farming | 869 |
| | | 01,6 | | Support activities to agriculture and post-harvest crop activities | |
| | | | 01,61 | Support activities for crop production | 0 |
| | | | 01,62 | Support activities for animal production | 109 |
| | | | 01,63 | Post-harvest crop activities | 19 |
| | | 01,7 | | Hunting, trapping and related service activities | 2 |
| | | | 01,70 | Hunting, trapping and related service activities | 5 |
| | 03 | | | **Fishing and aquaculture** | |
| | | 03,1 | | Fishing | |
| | | | 03,11 | Marine fishing | 227 |
| | | | 03,12 | Freshwater fishing | 0 |
| | | 03,2 | | Aquaculture | |
| | | | 03,21 | Marine aquaculture | 1 |
| | | | 03,22 | Freshwater aquaculture | 3 |

This approach would facilitate the decision-making processes related to the food waste, establishing a basic framework based on the number of entities linked with economic activities which are considerable numerous in relation to be potentially food wastage generators.

Thus, in the case of the Basque Country and with regard to the production section, a first approach before of measuring the food loss and waste would be to define the most numerous types of the economic activities with the potential to generate food wastage, as it is the case with the entities classified as *Mixed farming*, *Raising other cattle* and *Growing of grapes*, so that these types of entities, within the production sector, should be given higher priority in the establishment of food wastage measurements.

Furthermore, this methodology aims to avoid some inconsistencies related to the establishment of the scope to measure the food wastage using upper level categories instead of the class category because upper level categories could also include economic activities which are not linked to the potential generation of food wastage.

**Conclusions**

This document has sought to make a contribution by proposing a new methodology for facilitating the food wastage quantification with the aim of making progress on improving the food waste knowledge and evaluating the level of reliability of the current official figures at local and supra-local scale and it allows for the development of a comparative framework between different municipalities.

Thanks to the proposed methodology, it is possible to provide baseline information related to the number and the nature of the entities linked with economic activities with potential generation of food wastage and their geographical location in the territory. Thus, it could facilitate the work of the technician, administrations and policy-makers to establish strategies for the quantification of the food loss and waste, particularly with regard to the earliest steps of the measurement process to establish the scope of an FLW inventory.

Moreover, this methodology could represent a step forward in identifying and helping resolve some inconsistencies related to the inclusion of bio-waste data proceeding from economic activities not linked to the food wastage generation and to measure the food wastage issue in a certain territory. The opposite could also occur in the case of activities associated to food wastage production which are not included in the food waste figures.

Based on the identification of these gaps, it is possible to prioritise studies at local and regional scales so as to fill the existing lack of information regarding the food waste and improving the reliability of the official figures at different scales. Furthermore, the proposed methodology aims to facilitate the dialogue and discussion between different agents about the need to pave the way towards food waste quantification at the different levels of management (local, regional and national), particularly for searching a rigorous vision and diagnosis of the situation in order to establish a basis on which decision makers could define reduction targets in the short, medium and long term.

This report therefore emphasises the need to move towards more adjusted methodologies to quantify the food waste, using information that is already available. This would entail an effective and pragmatic way of helping create a diagnosis about the food loss and waste at local scale, but at the same time it seeks to provide a critical review to drive and lead new quantification studies about this problem at different scales. Thus, it would avoid remaining information with possible means of improvement as official figures because these data could not be used as the basis for carrying out strategies for the current food waste reduction. That aspect represents a fundamental step to address the problems and propose solutions or improvements of this global phenomenon which is having severe negative effects at economic, social, environmental levels as well as an important impact on social and ethical issues.

**Acknowledgement**
This work is part of the Waste4Think project. Waste4Think has received funding from a) the European Union's Horizon 2020 research and innovation programme under grant agreement 688995 and b) the Ph.D. granted by the University of Deusto (2016-2019) to Hector Barco. The authors would like to thank Dr. Oihane Kamara-Esteban for her valuable comments and suggestions and the town council of Zamudio and the Basque Statistical Office (EUSTAT) for sharing economical activities data. The constructive comments of two anonymous reviewers are also acknowledged.